\documentclass[twocolumn, pra, superscriptaddress]{revtex4-1}
\usepackage{amssymb}
\usepackage{amsmath}
\usepackage{epsfig}
\usepackage{color}
\usepackage{graphics, graphicx,subfigure}
\usepackage{bbold}
\usepackage{psfrag}
\usepackage{mathcomp}
\usepackage{subfigure}
\usepackage{verbatim}
\usepackage{color}
\usepackage{epstopdf}
\usepackage{lipsum}
\usepackage[colorlinks,citecolor=blue]{hyperref}
\def\cp#1{\mathbf{#1}}

\begin{document}
\title{Few-body solutions under spin-exchange interaction:\\
magnetic bound state and the Kondo screening effect}
\author{Cheng Peng}
\affiliation{Beijing National Laboratory for Condensed Matter Physics, Institute of Physics, Chinese Academy of Sciences, Beijing 100190, China}
\affiliation{School of Physical Sciences, University of Chinese Academy of Sciences, Beijing 100049, China}
\author{Xiaoling Cui}
\email{xlcui@iphy.ac.cn}
\affiliation{Beijing National Laboratory for Condensed Matter Physics, Institute of Physics, Chinese Academy of Sciences, Beijing 100190, China}
\affiliation{Songshan Lake Materials Laboratory, Dongguan, Guangdong 523808, China}
\date{\today}

\begin{abstract}
Motivated by recent progresses on ultracold alkaline-earth atoms towards the goal of simulating Kondo physics, in this work we exactly solve the few-body problem of one and two trapped fermions in one dimension interacting with a localized impurity under tunable spin-exchange interaction. 
It is found that depending on the sign of the spin-exchange coupling, ferromagnetic(FM) or anti-ferromagnetic(AFM), the attractive and repulsive branches can hold different magnetic structures. For the two fermions case, we demonstrate the  Kondo screening effect for the attractive branch with AFM coupling, and show that such screening  is absent for the ground state with FM coupling. Moreover,  we find a sequence of FM upper branches in the AFM coupling side. These FM states are orthogonal to all other attractive branches and their wave functions feature a full spin-charge separation. 
The effect of an additional contact interaction and the extension of our results to many particles are also discussed. This work reveals the intriguing physics uniquely associated with the spin-exchange interaction in the few-body point of view, which are promisingly to be explored in the experiment of ultracold alkaline-earth atoms.  
\end{abstract}

\maketitle

\section{Introduction}

Recent progresses on ultracold alkaline-earth atoms have opened the door for simulating Kondo physics\cite{Gorshkov}. In these two-electron atoms, the metastable excited state $^3P_0$ as well as the ground state $^{1}S_0$ comprise the two-orbital system, and the high nuclear spin of these atoms serves as the spin degree of freedom. The spin-exchange interaction between different orbitals, as a crucial ingredient for Kondo physics, has been confirmed experimentally\cite{Ludwig-Maximilians-2014,Fallani-2014,Fallani-2015,Ludwig-Maximilians-2015,JILA,Kyoto}.
%A fascinating property of these two-electron atoms is  naturally two-orbital systems, in that they not only have a stable ground state $$, but also a metastable electronically excited state, $$ or $$, which has a long lifetime to be accessible in experiments\cite{...}. Besides, the high nuclear spin of these atoms serves as the spin degree of freedom, and the spin-exchange interaction between different orbitals, which is a crucial ingredient for Kondo physics, has also been confirmed experimentally\cite{..}. 
The bare spin-exchange interaction is found to be ferromagnetic for $^{173}$Yb\cite{Ludwig-Maximilians-2014,Fallani-2014,Fallani-2015,Ludwig-Maximilians-2015} and $^{87}$Sr\cite{JILA}, and antiferromagnetic for $^{171}$Yb\cite{Kyoto}. Using the confinement-induced resonance, the strength of spin-exchange interaction can be conveniently tuned in the low dimensions, as successfully demonstrated both theoretically\cite{IASTU-2016,IASTU-2017,Zhang-Zhang-2018,Renmin-University,Zhang-Zhang-2020,Zhang-Cheng-Zhang-Zhai} and experimentally\cite{Ludwig-Maximilians-2018}. Furthermore, the Kondo model requires a local impurity, which can be implemented in alkaline-earth atoms by using a proper laser wavelength to selectively confine $^3P_0$ state while let $^{1}S_0$ state free\cite{NIST,Ludwig-Maximilians-2018}.  All these developments make the quantum simulation of Kondo physics quite promising in ultracold atomic systems.   

Motivated by these developments, in this work we exactly solve the problems of a few  trapped fermions in one-dimension (1D) interacting with a local impurity under the spin-exchange interaction. Specifically, we consider the isotropic Heisenberg coupling  between fermions and the impurity with a tunable coupling strength $J$. 
Early in the 1980's, the 1D Kondo model describing the continuum fermions and a  magnetic impurity was exactly solved by Bethe-ansatz method assuming a linear dispersion of fermions($\epsilon_k\propto k$)\cite{Andrei-Furuya-Lowenstein}. 
Given this assumption, the results of Ref.\cite{Andrei-Furuya-Lowenstein} are  expected to be applicable in weak coupling limit, where only the fermions near the Fermi surface are perturbed by the impurity scattering. In comparison, in this work we aim at exactly solving a few trapped fermions interacting with a local impurity for an {\it arbitrary} coupling strength $J$, in particular, including the strong coupling regime which is not covered by Ref.\cite{Andrei-Furuya-Lowenstein}.

In this work, we reveal a number of unique properties associated with spin-exchange interaction, which cannot be achieved by the pure contact potential without spin exchange. For instance, depending on the sign of spin-exchange coupling $J$, namely, $J<0$ for ferromagnetic(FM) coupling  or $J>0$ for anti-ferromagnetic(AFM) coupling, it is found that the attractive and repulsive branches of the system can hold different magnetic structures. Importantly, for the attractive branches in the two fermions case, we demonstrate the Kondo screening effect under the AFM coupling, and show that such screening does not apply to the case of FM coupling.
%Specifically, the ferromagnetic (anti-ferromagnetic) coupling would support triplet (singlet) -type bound states. 
Moreover,  we find a sequence of FM upper branches in the AFM coupling side, which are orthogonal to all other attractive branches and feature a full spin-charge separation. 
All these features are closely related to the spin-exchange nature of the interaction potential, and can be readily probed in the experiments of alkali-earth atoms. %Moreover, we derive the analytical expression of  energy spectrum for different branches in both weak and strong coupling regimes, and finally discuss its response to an additional contact interaction between fermions and impurity. 
The effect of an additional contact interaction and the extension of our results to many particles are also discussed in this work.

In section \ref{sec2}, we will present the formula of exactly solving one and two fermions interacting with the impurity through the spin-exchange coupling. The few-body results as well as the extension to many particles are presented in section \ref{sec3}. We further discuss the effect of an additional contact potential in section \ref{sec4} and the experimental relevance of our results in section \ref{sec5}. Finally, we conclude our work in section \ref{sec6}.  

\section{Exact formula of few-body problem with spin-exchange interaction} \label{sec2}

\subsection{Model}
We consider the 1D harmonically trapped spin-$1/2$ fermions interacting with a local spin impurity (sitting at the trap center) with a spin-exchange coupling, which is described by Hamiltonian ($\hbar=1$ throughout the paper):
\begin{eqnarray}
H&=&H_0+U, \label{H} \\
H_0&=& \sum_{i=1}^{N} \left( -\frac{1}{2M} \frac{\partial^2}{\partial x_{i}^2}   +\frac{M\omega^2}{2}x_{i}^2 \right); \\
U&=& 2J\sum_{i=1}^N\delta(x_i)\bf{S}_i\cdot \bf{S} 
\end{eqnarray}
Here $M$ is the fermion mass; $x_i$ is the coordinate of the $i$-th fermion; ${\bf S}_i=(S_{ix},S_{iy},S_{iz})$ and ${\bf S}=(S_x,S_y,S_z)$ are the spin operators for, respectively, the $i$-th fermion and the impurity. 
The spin operator of fermions can be expressed by the fermion field operators as
\begin{eqnarray}
S_{ix}&=&\frac{1}{2}(\psi_{\uparrow}^{\dag}(x_i)\psi_{\downarrow}(x_i)+\psi_{\downarrow}^{\dag}(x_i)\psi_{\uparrow}(x_i));\nonumber\\
S_{iy}&=&\frac{-i}{2}(\psi_{\uparrow}^{\dag}(x_i)\psi_{\downarrow}(x_i)-\psi_{\downarrow}^{\dag}(x_i)\psi_{\uparrow}(x_i));\nonumber\\
S_{iz}&=&\frac{1}{2}(\psi_{\uparrow}^{\dag}(x_i)\psi_{\uparrow}(x_i)-\psi_{\downarrow}^{\dag}(x_i)\psi_{\downarrow}(x_i)).
\end{eqnarray}
Here $\psi_{\sigma}^{\dag}(x)$ is to create a spin-$\sigma(=\uparrow,\downarrow)$ fermion at coordinate $x$, which can be further expanded as
\begin{equation}
\psi_{\sigma}^{\dag}(x)=\sum_m C_{m\sigma}^{\dag} \phi_m(x),
\end{equation} 
with $C_{m\sigma}^\dagger$  the creation operator of a spin-$\sigma$ fermion at the $m$-th harmonic oscillator level with eigen-wavefunction $\phi_m(x)$ and eigen-energy $E_m = (m+\frac{1}{2})\omega$. In this way, the Hamiltonian (\ref{H}) can be written in the second quantized form:
\begin{equation}
\begin{split}
H &= \sum_{m\sigma}E_m C_{m\sigma}^\dagger C_{m\sigma} + \sum_{m,n} V_{mn} ( C_{m\uparrow}^\dagger  C_{n\downarrow} S_- + h.c. \\
     & +   (C_{m\uparrow}^\dagger  C_{n\uparrow}-C_{m\downarrow}^\dagger  C_{n\downarrow}) S_z ), \label{H2}\\
\end{split}
\end{equation}
with $V_{mn} = J\phi_m(0)\phi_n(0)$ the coupling matrix element generated by the spin-exchange interaction $U$. Denoting $|\Uparrow\rangle$ and $|\Downarrow\rangle$ as the two spin states of the impurity, we then have $S_+=|\Uparrow\rangle\langle \Downarrow|$,  $S_-=|\Downarrow\rangle\langle \Uparrow|$ , and $S_z=(|\Uparrow\rangle\langle \Uparrow|-|\Downarrow\rangle\langle \Downarrow|)/2$.  
One can see clearly that the spin-exchange process is governed by the first two terms in the bracket of (\ref{H2}). 

Next, we will present the formula of exactly solving the few-body problems of one or two fermions interacting with a local impurity through spin-exchange coupling.

\subsection{One fermion plus an impurity}

Because the Hamiltonian (\ref{H2}) preserves the spin-rotational symmetry, the one-fermion eigenstate can be classified into two cases with respect to the total spin of the fermion and the impurity, denoted by  ${\bf S}_{tot}$. One case is spin triplet with ${\bf S}_{tot}=1$, the other case is spin singlet with ${\bf S}_{tot}=0$. We then have the effective interaction for each spin channel:
\begin{equation}
U_{s,t}(x)=\gamma_{s,t}\delta(x)
\end{equation}
with effective coupling strengths $\gamma_s = \frac{-3J}{2}$ and $\gamma_t = \frac{J}{2}$, respectively, for spin-singlet and spin-triplet channels.  

In order to cover both spin-singlet and triplet states in the same ansatz, we consider the total zero magnetization ($S_{tot,z}=0$) subspace and write down the ansatz as
\begin{equation}
|\Psi\rangle_2=\sum_m \left( \phi_m^{(1)} C_{m\uparrow}^\dagger|0\rangle |\Downarrow\rangle+ \phi_m^{(2)} C_{m\downarrow}^\dagger|0\rangle |\Uparrow\rangle \right)
\end{equation}
Here $|0\rangle$ is the vacuum state of fermion. By imposing the Schrodinger equation $H |\Psi \rangle_2 = E |\Psi\rangle_2$, we arrive at the following coupled equations: 
    \begin{equation}
    \begin{split}
      (E-E_m) \phi^{(1)}_m &= \sum_p (V_{mp}\phi^{(2)}_p-\frac{1}{2}V_{mp}\phi^{(1)}_p )\\
      (E-E_m) \phi^{(2)}_m &= \sum_p (V_{mp}\phi_p^{(1)} -\frac{1}{2}V_{mp}\phi^{(2)}_p)\\
    \end{split}
    \end{equation}
We can see that above equations support two types of solutions. One is     \begin{equation}
      \phi^{(1)}_m=\phi^{(2)}_m \propto \frac{\phi_m(0)}{E-E_m},
    \end{equation}
which represents a spin triplet, and the energy $E(=E_t)$ can be obtained from     \begin{equation}
      \frac{1}{\gamma_t} =  \sum_m \frac{|\phi_m(0)|^2}{E_t-E_m}. \label{eq_t}
    \end{equation}
The other is spin singlet solution with 
    \begin{equation}
      \phi^{(1)}_m = -\phi^{(2)}_m\propto \frac{\phi_m(0)}{E_s-E_m},
    \end{equation}
and  the energy $E(=E_s)$ follows:
    \begin{equation}
      \frac{1}{\gamma_s} =  \sum_m \frac{|\phi_m(0)|^2}{E_s-E_m}. \label{eq_s}
    \end{equation}
Eqs.(\ref{eq_t},\ref{eq_s}) can be further simplified as 
    \begin{equation}
        -\frac{2\sqrt{\pi}}{\kappa_{s,t}} = B(-\frac{\rho_{s,t}}{2},\frac{1}{2}) 
    \end{equation}
    with $\kappa_{s,t} \equiv \gamma_{s,t}\sqrt{M/\omega}$, $\rho_{s,t} \equiv E_{s,t}/\omega-1/2$ and $B(x,y)$ is the beta function.

\subsection{Two fermions plus an impurity}

For the few-body system consisting of two fermions and the impurity, we can have the total spin ${\cp S}_{tot}=3/2$ or ${\cp S}_{tot}=1/2$. Again in order to cover both states in the same ansatz, we consider the subspace with total magnetization $S_{tot,z}=1/2$ and write down the ansatz as
   \begin{equation}
    |\Psi\rangle_3 = \sum_{mn} \left( \phi^{(1)}_{mn} C_{m\uparrow}^\dagger C_{n\uparrow}^\dagger \left|0\right> |\Downarrow\rangle + \phi^{(2)}_{mn}  C_{m\uparrow}^\dagger C_{n\downarrow}^\dagger \left|0\right> |\Uparrow\rangle \right)
   \end{equation}
Here we should take care of the anti-symmetry property of $\phi^{(1)}_{mn}$ , i.e., $\phi^{(1)}_{mn} = -\phi^{(1)}_{nm}$. Again by imposing the Schrodinger equation, we obtain the following coupled equations:
\begin{widetext}
    \begin{equation}
        \begin{split}
        &\phi^{(1)}_{mn} = \frac{1}{E-E_m-E_n} \cdot \frac{1}{2} \cdot  \sum_{p}-V_{mp}\phi^{(2)}_{np}+V_{np}\phi^{(2)}_{mp} +V_{np}\phi^{(1)}_{pm}- V_{mp}\phi^{(1)}_{pn} \\
        &\phi^{(2)}_{mn} = \frac{1}{E-E_m-E_n} \sum_p V_{np}\phi^{(1)}_{mp}-V_{np}\phi^{(1)}_{pm}+ \frac{1}{2} V_{mp}\phi^{(2)}_{pn}- \frac{1}{2}V_{np}\phi^{(2)}_{mp}\\
        \end{split}\label{eq_3b}
    \end{equation} 
\end{widetext}
 %  	with $V_{mp} = J \phi_m\phi_p $.
To solve these coupled equations, we introduce three series of variables $F^{(1)}_m,F^{(2)}_m,F^{(3)}_m$  :
    \begin{equation}
        \begin{split}
        &F^{(1)}_n \equiv \sum_p\phi_p(0)\phi^{(1)}_{np} \\
        &F^{(2)}_n \equiv \sum_p\phi_p(0)\phi^{(2)}_{np} \\
        &F^{(3)}_n \equiv -\sum_p\phi_p(0)\phi^{(2)}_{pn}\\
        \end{split} \label{F}
    \end{equation}
Then we can multiply both sides of (\ref{eq_3b}) by $\phi_m(0)$ and sum over $m$ to obtain the coupled equations of $F^{(i)}_m$. To see more clearly the physical meaning of the variables in (\ref{F}), we can alternatively perform a linear transformation of them to a different set of variables:
    \begin{equation}
        \begin{split}
        &\tilde{F}^{(1)}_n =-\frac{3}{2}F_n^{(1)} +\frac{3}{4}F_n^{(2)}\\
        &\tilde{F}^{(2)}_n =\frac{1}{2}F_n^{(1)}+\frac{1}{4}F_n^{(2)} \\
        &\tilde{F}^{(3)}_n =\frac{1}{2}F_n^{(3)}\\
        \end{split} \label{tF}
    \end{equation}
 Then we find that $\{ \tilde{F}^{(i=1,2,3)} \}$ are exactly the atom-dimer amplitudes in the wave function:   
 \begin{equation}
    \begin{split}
        |\Psi\rangle_3 &=\sum_m \tilde{F}^{(1)}_m \left| m \uparrow \right> \left|d^{00}_{m}\right> + \tilde{F}^{(2)}_m \left|m\uparrow \right>  \left| d^{10}_{m}\right> +   \tilde{F}^{(3)}_m \left|m\downarrow\right> \left|d^{11}_{m}\right>\\
    \end{split}
    \end{equation}
 with the dimer states:   
  \begin{eqnarray}
    \left|d^{11}_{m}\right> &=& \sum_p \frac{\phi_p(0)}{E-E_m-E_p} \left| p\uparrow \right>|\Uparrow\rangle ; \nonumber\\
    \left|d^{10}_{m}\right> &=& \sum_p \frac{\phi_p(0)}{E-E_m-E_p} \frac{\left| p\uparrow \right>|\Downarrow\rangle+\left| p\downarrow \right>|\Uparrow\rangle}{\sqrt{2}} ; \nonumber\\
\left|d^{00}_{m}\right> &=& \sum_p \frac{\phi_p(0)}{E-E_m-E_p} \frac{\left| p\uparrow \right>|\Downarrow\rangle-\left| p\downarrow \right>|\Uparrow\rangle}{\sqrt{2}} . \nonumber
  \end{eqnarray}

Denoting $\tilde{F}^{(i)}\equiv (\tilde{F}^{(i)}_0,\tilde{F}^{(i)}_1,...)^T$, finally we arrive at the following matrix equation from (\ref{eq_3b}):  
     \begin{equation}
      \left(
        \begin{array}{ccc}
        \frac{1}{4} (e-2 q) 3 & \frac{3}{4} e  & \frac{-3}{4}  e  \\
        -\frac{1}{4}e & -\frac{1}{4} (e-2 q) & -\frac{1}{4} e  \\
        \frac{1}{2} e  & -\frac{1}{2} e  & \frac{1}{2} q  \\
        \end{array}
      \right)
        \left(
            \begin{array}{c}
                \tilde{F}^{(1)} \\
                \tilde{F}^{(2)} \\
                \tilde{F}^{(3)} \\
            \end{array}
        \right)
        =\frac{1}{J}
        \left(
            \begin{array}{c}
                \tilde{F}^{(1)} \\
                \tilde{F}^{(2)} \\
                \tilde{F}^{(3)} \\
            \end{array}
        \right)   \label{final_eq}
    \end{equation}   
where e and q are all matrixes with elements $e_{mn} = \frac{\phi_m(0) \phi_n(0)}{E-E_m-E_n}$ and $q_{mn} = \delta_{mn}  \sum_p \frac{|\phi_p(0)|^2}{E-E_m-E_p}$. The $q$-matrix is due to the interaction between one fermion and the impurity (forming a dimer), while $e$-matrix is due to interaction between the dimer and the other fermion. In practical simulation, we find convergent results can be reached with cutoff $N_{c}=100$ for the element indices of $e$- and $q$-matrixes.  Accordingly, (\ref{final_eq}) is a $3N_{c}\times 3N_{c}$ matrix.

\section{Results} \label{sec3}

In this section, we present the results of one fermion and two fermions cases based on the formula shown in previous section. Given the even parity of interaction potential ($x\leftrightarrow-x$), we only present the few-body results associated with even-parity wave functions but neglect the odd-parity ones which are not affected by the interaction. 

\subsection{One fermion plus an impurity}

\begin{figure}[h]
\includegraphics[width=8cm]{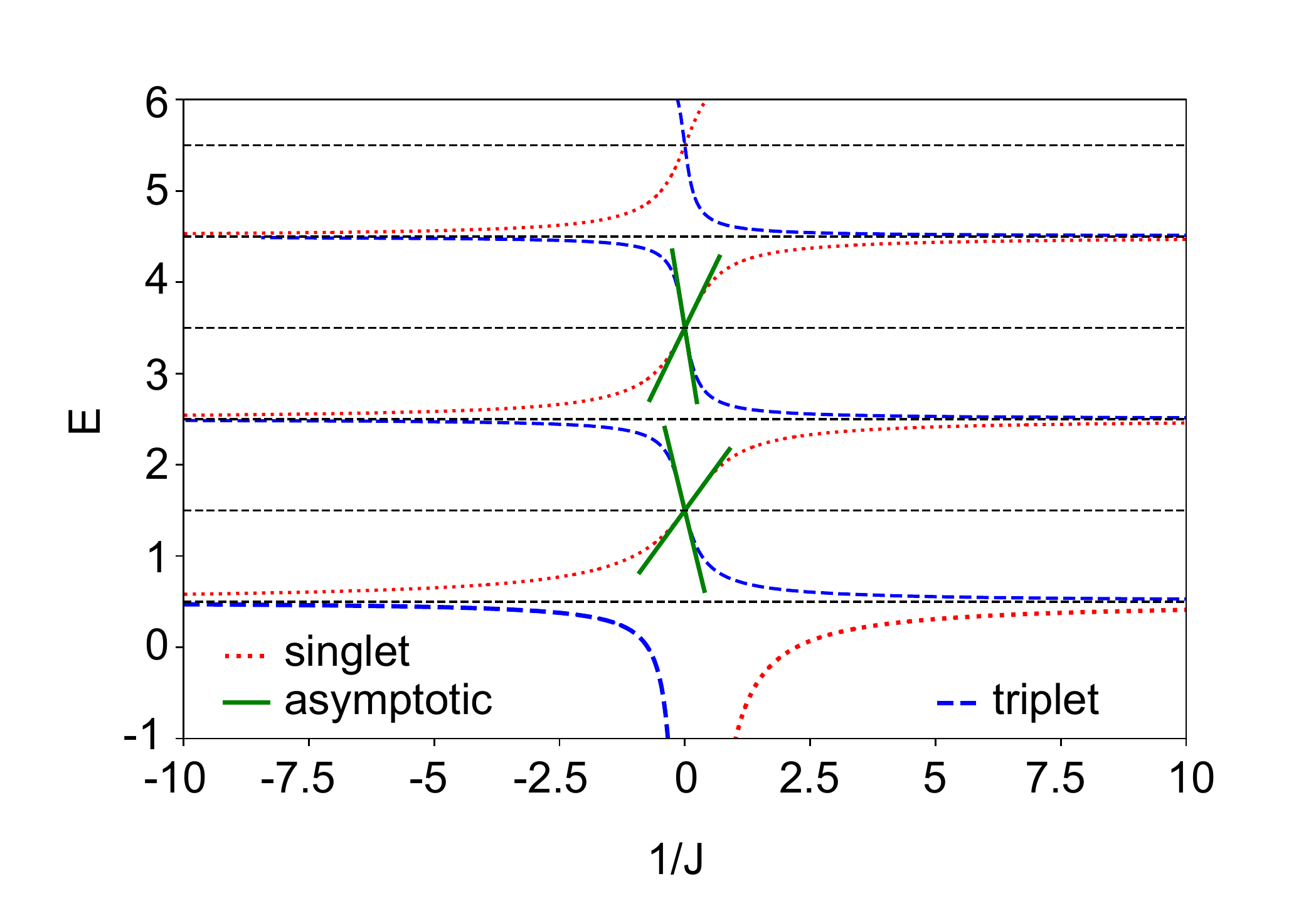}
\caption{(Color online). Energy spectrum of one fermion and one impurity in $S_{tot,z}=0$ subspace. The blue dashed (red dotted) lines show $E_t$ ($E_s$) for spin-triplet (singlet) eigen-states. The green solid line shows the asymptotic fitting to (\ref{asymptotic}) in strong coupling limit. Here the units of $E$ and $J$ are respectively $\omega$ and $\sqrt{\omega/M}$. 
} \label{fig_2body}
\end{figure}

In Fig.\ref{fig_2body}, we show the energy spectra for both spin-triplet ($E_t$) and spin-singlet ($E_s$) states, obtained by solving, respectively, (\ref{eq_t}) and (\ref{eq_s}). 

For the spin-singlet case, as the effective coupling is given by $\gamma_s=-3J/2$, therefore the bound state below the scattering threshold will be supported at $J>0$. As shown in Fig.\ref{fig_2body}, as increasing $J$ from zero to $\infty$, the singlet bound state becomes deeper and deeper, and the energy approaches $E_s\rightarrow -9J^2/8$ in large $J$ limit. This is the so-called lower attractive branches. Meanwhile, we also see another sequence of eigenstates with energies approaching finite values when $J\rightarrow\infty$. These branches are well above the attractive branch in the spectrum, therefore called the repulsive upper branches. For this type of energy branches, one can carry out an effective perturbation theory in terms of small value $1/J\rightarrow 0$. As shown in Appendix \ref{app_a1}, the $n$-th upper branch energy approaches
\begin{equation}
E_{s,n} \simeq (2n+1)\omega -\frac{\phi^{'}_{2n+1}(0)^2}{M^2\gamma_s} , \label{asymptotic}
\end{equation}
here $n$ is the level index, and $\phi'_n=d\phi_n(x)/dx$ is the first derivative of single-particle state $\phi_n$. Accordingly, the zero-th order wave function is\cite{Heidelberg}:
\begin{equation}
|\Psi\rangle_{s,n} = \phi_{2n+1}(x) {\rm sgn}(x)\left| \uparrow\Downarrow- \downarrow\Uparrow\right>, \label{asymptotic_wf}
\end{equation}
with the sign function ${\rm sgn}(x)=x/|x|$, which is $1\ (-1)$ for $x>0\ (<0)$.

For the spin-triplet state, the analysis is similar, except now the effective coupling is $\gamma_t=J/2$. Therefore the spin-triplet bound state below the scattering threshold is supported at $J<0$ side, with the binding energy approaches $E_t\rightarrow -J^2/8$ in large $|J|$ limit. The asymptotic behavior of upper branches follows the form of (\ref{asymptotic}) by replacing $E_s,\gamma_s$ with $E_t,\gamma_t$. The according zero-th order wave function follows the spin-charge separation form of (\ref{asymptotic_wf}) with the spin part replaced by $\left| \uparrow\Downarrow+ \downarrow\Uparrow\right>$.

One thus can summarize that for the one fermion case, once considering the effective coupling for each spin channel, the resulted spectrum is quite similar to the case of a pure contact potential, as solved originally by T. Busch {\it et al}\cite{Busch-Englert-Rzazewski-Wilkens}. However, for the two fermions case, we will show below that the resulted spectrum is very rich due to the spin-exchange interaction, which is very  different from the case of a pure contact potential.

\subsection{Two fermions plus an impurity}

\begin{figure}[t]
\includegraphics[width=7cm]{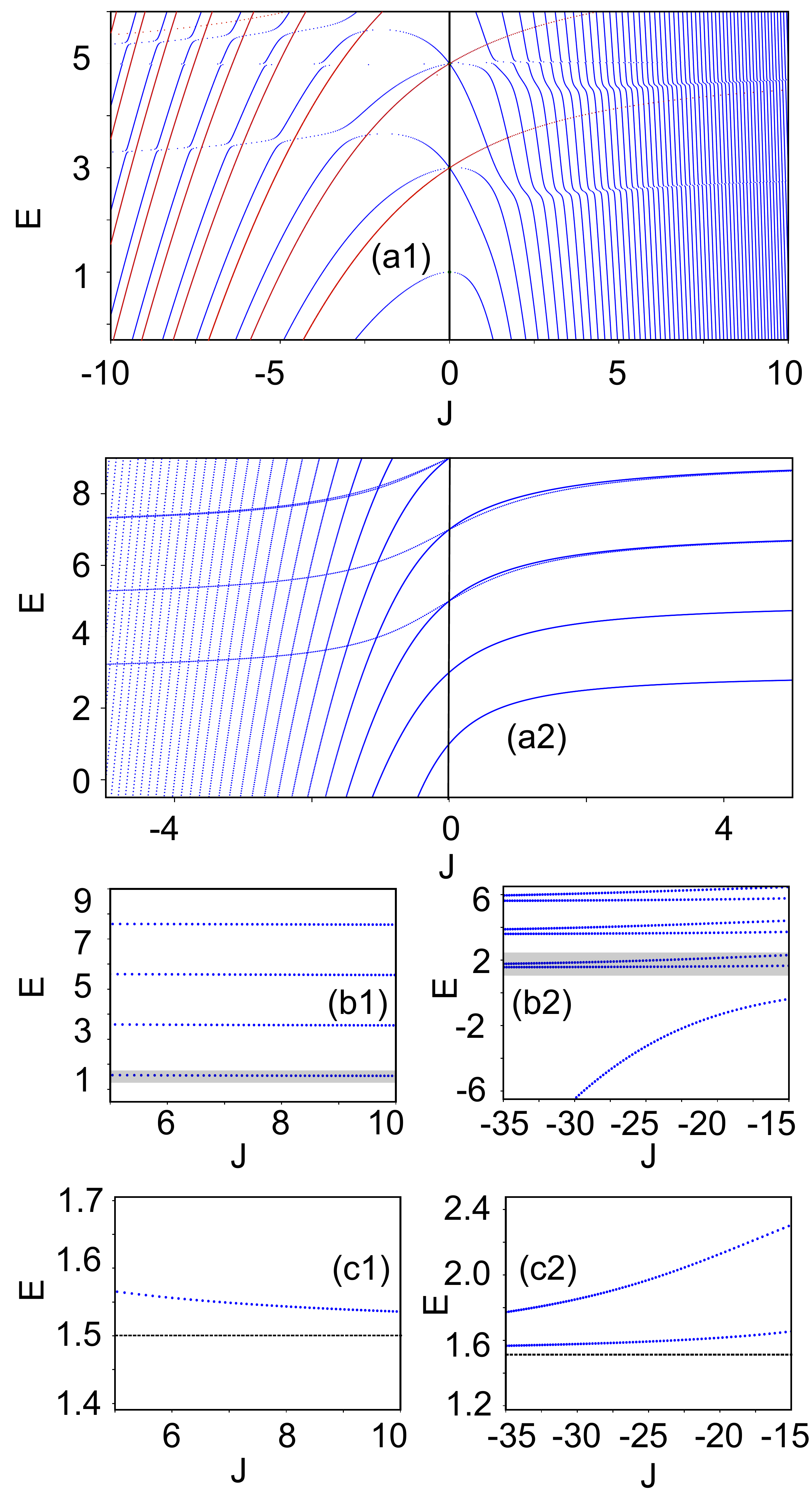}
\caption{(Color online).(a1) Energy spectrum of two fermions and one impurity with spin-exchange interaction in the $S_{tot,z}=1/2$ subspace. The FM states with $S_{tot}=3/2$ are highlighted by red color, and the rest are all with total spin $S_{tot}=1/2$. (a2) Same as (a1) except that the interaction between fermions and impurity is a pure contact type without spin-exchange, see text. (b1) Energies of deep bound states  for large and positive $J$, shifted by the spin-singlet binding energy $-9J^2/8$. All shifted energies saturate to a finite value, signifying the Kondo screening effect (see text).(b2) Energies of deep bound states  for large and negative $J$, shifted by the spin-triplet binding energy $-J^2/8$. The lowest bound state does not saturate as in (b1), showing the absence of screening effect for FM coupling. (c1,2) show the magnified plot for the shaded region in (b1,2), illustrating the asymptotic behavior towards odd harmonic levels.
The units of $E$ and $J$ are respectively $\omega$ and $\sqrt{\omega/M}$. 
%{\color{red} Put/merge (a,b,c) in a single pdf file. The visibility of labels, curves, and numbers in these figures still need to improve, especially b and c!  Remove all insets from b and c.)}
} \label{fig_3body}
\end{figure}

In Fig.\ref{fig_3body}, we show the energy spectrum of the system consisting two fermions and an impurity in the $S_{tot,z}=1/2$ subspace. In particular, we mark the FM state (with the largest total spin $S_{tot}=3/2$) with red colors, and the rest are all with $S_{tot}=1/2$.

Compared to the system with a pure contact interaction, we can see from Fig.\ref{fig_3body} that the current system with spin-exchange interaction displays very rich spectrum. Here we would like to highlight three properties that are uniquely associated with the spin-exchange interaction. 

First, many branches of bound states below the scattering threshold appear in both $J>0$ and $J<0$ sides(see Fig.\ref{fig_3body}(a1)), which are associated with different magnetic structures.  This is very different from the case of pure contact interaction without spin exchange, i.e., $U=2J\sum_i \delta(x_i)$. As shown in Fig.\ref{fig_3body}(a2), in the pure contact $U$ case,  the bound states can only appear in $J<0$ side.  In comparison, the spin-exchange interaction can lead to much richer spectrum. Moreover, we can see that for large and positive $J$, the bound state energies are dominated by the energy of spin singlet (composed by the impurity and one fermion, with energy $\sim -9J^2/8$), and the residue interaction between the spin-singlet and the other fermion only contributes constant energies (e.g., $1.5\omega, 3.5\omega...$, see Fig.\ref{fig_3body}(b1,c1)), which are much smaller than the spin singlet binding energy. Therefore, the impurity appears to be nearly screened through the formation of spin-singlet bound state with the first fermion, which cannot further attract the other fermion. This resembles the Kondo screening effect in metals, where the magnetic spin of the localized impurity is screened by forming a spin singlet with conduction electrons\cite{book_Kondo}. On the contrary,  the lowest bound state in $J<0$ side is not simply dominated by a single spin-triplet bound state; rather, the impurity can attract both fermions to produce a deeper bound state with binding energy $E\approx -0.151J^2$($<E_t=-J^2/8$), see Fig.\ref{fig_3body}(b2)). We have also checked this deeply bound state for two untrapped fermions interacting with a localized impurity, which produces the same binding energy as the trapped fermions case in the limit of large $|J|$ (for more details see Appendix \ref{app_d}). 
This means that the impurity cannot be screened by forming a spin-triplet with a fermion, in contrast with the AFM coupling case. This is because the spin-triplet bound state is magnetic, and the other fermion can still interact with it to form an even deeply bound state as ground state. The situation is different for the higher excited branches. As shown in Fig.\ref{fig_3body}(c2), the energies of these excited branches are still dominated by the spin-triplet binding energy.

Second, a sequence of full FM states(with the largest total spin ${\bf S}_{tot}=3/2$) can be found in upper branches of the AFM coupling ($J>0$) side, as marked by red color in the spectrum in Fig.\ref{fig_3body}(a1). Since all attractive branches in $J>0$ side contain a spin-singlet part (formed by one fermion and the impurity), they all have the total spin ${\bf S}_{tot}=1/2$ and thus are all orthogonal to the FM branches (with ${\bf S}_{tot}=3/2$). Therefore, once an initial state is prepared in the FM branch, it will always stay on this branch due to zero-coupling with other branches. Another interesting feature of these FM branches is that their wave functions are fully spin-charge separated. This is because in a full FM state, any two particles will form a triplet pair. Therefore, the interaction potential can be simplified as $U_t(x_i)=\gamma_t\delta(x_i)$, which is only relevant to the charge part. As a result, one can construct the FM wave function of two-fermion system as:
\begin{equation}
\Psi_{3,FM}(x_1,x_2)=\left|\begin{array}{cc}\psi_1(x_1) & \psi_1(x_2) \\\psi_2(x_1) & \psi_2(x_2)\end{array}\right| \left(|\uparrow\downarrow\Uparrow\rangle+|\downarrow\uparrow\Uparrow\rangle+|\uparrow\uparrow\Downarrow\rangle\right),
\end{equation}
where $\psi_{1,2}$ is any of the triplet eigenstates for the system of one fermion plus the impurity, and the spin part is a full FM state. One can see clearly this wave function is an eigen-state of the Hamiltonian, with eigen-energy $E=E_1+E_2$ ($E_{1,2}$ is the eigen-energy of $\psi_{1,2}$ for one-fermion system). 
Note that these FM branches should be distinguished from that in the 1D spin-1/2 fermion system\cite{Cui-Ho}, where the FM state do not feel any s-wave interaction and the energy are always static as changing coupling strength.
The asymptotic expansion of the energies of FM states and the other branches will be presented in Appendix \ref{app_a2}. 

Third, in the weak coupling limit $J\rightarrow 0^{\pm}$, we see that the ground state energy of the two fermions system exhibits a quadratic scaling as $E(J)\sim -cJ^2$ (see Fig.\ref{fig_3body}(a1)), instead of a linear one in the case of contact potential (see Fig.\ref{fig_3body}(a2)). This is because in the non-interaction limit, the two fermions form a spin singlet with the same orbital wave function (at the lowest harmonic oscillator level). It then follows that the expectation value of the spin-exchange interaction in this state is zero. This means that there is no mean-field contribution to the interaction energy, and the lowest level of energy correction comes from the second-order process (leading to $E\propto J^2$), which involves excitations to higher orbitals (i.e., higher harmonic oscillator levels). Note that this is different from the one-fermion case, where there is only one fermion and the mean-field contribution does exist and give the energy linearly depending on $J$ in weak coupling limit.

 We note that the energy scaling here is qualitatively different from that given by  the Beta Ansatz(BA) solution of continuum fermion case by assuming a linear dispersion of fermions\cite{Andrei-Furuya-Lowenstein}. From the BA study, we can extract a cubic energy scaling $E\propto J^3$, see Appendix \ref{app_c}. In comparison, we have shown a linear scaling ($E\propto J$) for one fermion and a quadratic scaling ($E\propto J^2$) for two fermions case in the weak coupling limit. The differences could be attributed to the presence of trapping potential in our case, which leads to discrete energy levels instead of continuous ones. As a result, the weak coupling regime in our case refers to $|J|\ll \sqrt{\omega/M}$  (where $\omega$  is the trapping frequency), while in Ref.\cite{Andrei-Furuya-Lowenstein} refers to $|J|\ll \sqrt{E_F/M}$  (where $E_F$ is the Fermi energy). In other words, the weak coupling regime in our case requires the spin-exchange energy much smaller than the single-particle gap, while in Ref.\cite{Andrei-Furuya-Lowenstein} requires it much smaller than the Fermi energy.  Therefore these two cases cannot be directly compared with each other. 

Finally, we would like to remark here that the three features discussed above, including the Kondo screening effect, the coexistence of FM upper branches and attractive lower branches, and the quadratic energy scaling, are all closely related to the spin-exchange nature of the interaction, which cannot exist in the system with pure contact interaction. These features can be generalized to many-fermion system as discussed below.

\subsection{Extension to many-body}

The one fermion and two fermions solutions have provided us an important insight to the properties of many-body system, where many spin-$1/2$ fermions (with number $N$) interact with a localized impurity with spin-exchange coupling $J$. These properties include:

First, the Kondo screening effect is expected to be applicable to the lowest bound state in the AFM coupling($J>0$) side, which tells that the impurity is essentially screened by forming a singlet bound state with one fermion (at the Fermi surface). Indeed, we note that in literature, a simple variational ansatz, which assumed a spin-singlet bound state on top of a unperturbed Fermi sea, was employed to estimate the Kondo temperature in metals\cite{book_Kondo}. On the other hand, our two-fermion calculation also suggests a finite residue interaction energy between the rest fermions and the singlet bound state, so they are not completely independent. How this residue interaction affects the Fermi sea atoms and the Kondo physics surely needs further investigation, which are beyond the scope of this work. 

Second,  a sequence of FM upper branches with total spin ${\bf S}_{tot}=(N+1)/2$ should exist in the AFM coupling($J>0$) side, which are orthogonal to all other branches in this regime. The wave functions of FM states feature a full spin-charge separation:
\begin{equation}
\Psi_{N+1,FM}(x_1,x_2,...x_N)=\left|\begin{array}{cccc}\psi_1(x_1) & \psi_1(x_2) & ... &\psi_1(x_N)  \\ \psi_2(x_1) & \psi_2(x_2) & ... &\psi_1(x_N) \\
... & ... & ... &... \\
\psi_N(x_1) & \psi_N(x_2) & ... &\psi_N(x_N)\end{array}\right| \left|FM\right>,
\end{equation}
where $\psi_{i}$ is a triplet eigenstate for the one-fermion system with eigen-energy $E_i$, and $\left|FM\right>$ is the FM spin state with ${\bf S}_{tot}=(N+1)/2$ and a specific $S_z$.  The total energy of $\Psi_{N+1,FM}$ is $E=\sum_{i=1}^N E_i$.

Third, the coupling dependence of ground state energy in the weak coupling limit $J\rightarrow 0^{\pm}$ will depend on whether the fermion number $N$ is even or odd. When $N$ is even, then in non-interacting limit the fermions are composed of $N/2$ pairs of spin singlet, and there is no mean-field contribution because the expectation value of the spin-exchange interaction in this state is zero. In this case, $E$ scales quadratically as $E(J)\sim J^2$. When $N$ is odd, then the single fermion at the Fermi surface will contribute to the mean-field energy as $E(J)\sim J$. However, in the thermodynamic limit ($N\rightarrow\infty$), the energy per particle $E/N$ will be dominated by quadratic term $\sim J^2$ in $J\rightarrow 0$ regime regardless of even or odd $N$, because the coefficient of linear dependence (for odd $N$) approaches zero when divided by $N$. This is different from the pure contact interaction case, where the energy per particle is always dominated by the mean-field contribution ($\sim J$) in weak coupling regime.

\section{Effect of an additional contact interaction} \label{sec4}

In this section, we discuss the effect of an additional contact interaction between fermion and the impurity, $U(x)=U\delta(x)$. The full Hamiltonian then reads
\begin{equation}
\begin{split}
     \hat{H} &= \sum_{m\sigma}E_m C_{m\sigma}^\dagger C_{m\sigma} + \sum_{m,n} V_{mn} ( C_{m\uparrow}^\dagger  C_{n\downarrow} S^- + h.c. \\
     & +   (C_{m\uparrow}^\dagger  C_{n\uparrow}-C_{m\downarrow}^\dagger  C_{n\downarrow}) S_z )+U_{mn} (C_{m\uparrow}^\dagger  C_{n\uparrow}+C_{m\downarrow}^\dagger  C_{n\downarrow})\\
\end{split}  \label{H_U}
\end{equation}
with $U_{mn}=U\phi_m(0)\phi_n(0)$.

\begin{figure}[h]
\includegraphics[width=6cm]{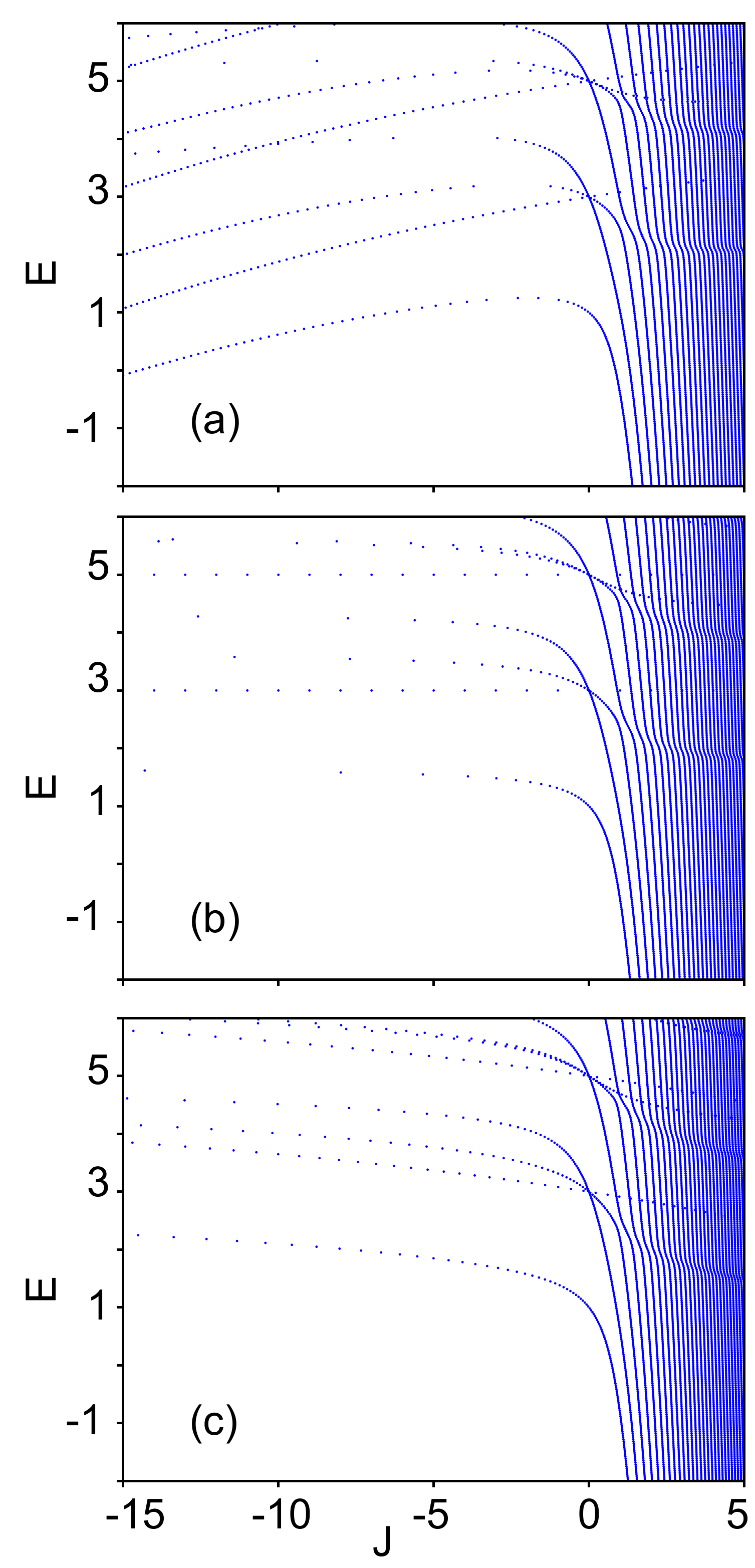}
\caption{(Color online).  Energy spectrum of two fermions and one impurity with spin exchange interaction and an additional contact interaction. We take three $\frac{U}{J} = -0.4,-0.5,-0.6$, respectively, in (a), (b) and (c). The units of $E$ and $J$ are respectively $\omega$ and $\sqrt{\omega/M}$.
%{\color{red} Again put (a,b,c) in a single file.} 
} 
\label{fig_U}
\end{figure}

The effect of contact interaction can be seen clearly from the one fermion case, where the effective interactions in the singlet and triplet channels are respectively given by 
\begin{equation} 
\gamma_s=-3J/2+U,\ \ \ \ \gamma_t=J/2+U. \label{eff_gamma}
\end{equation}
Therefore, the spin-singlet bound states are supported at $J>2U/3$ ($\gamma_s<0$), and the FM bound states are supported at $J<-2U$ ($\gamma_t<0$). 

The bound states of two-fermion system also have similar properties, in that the presence of contact $U$ will change the parameter regime to support ${\bf S}_{tot}=1/2$ or ${\bf S}_{tot}=3/2$(FM) bound states. The formula of solving two-fermion problem based on the Hamiltonian (\ref{H_U}) is given in Appendix \ref{app_b}. In  Fig.\ref{fig_U} we show the spectrum for three typical values of $U/J$. One can see that as changing $U/J$ across $-0.5$, the original bound states in $J<0$ regime gradually vanish since the effective FM coupling strength $\gamma_t$ changes from negative to positive. Similarly, we expect the vanishing of ${\bf S}_{tot}=1/2$ bound states (originally in $J>0$ regime) will occur when increasing the contact interaction to $U>3J/2$. The experimental relevance of our results will be discussed in the next Section.

\section{Experimental relevance} \label{sec5}

In this section, we discuss the experimental relevance of our model and results using ultracold alkaline-earth atoms. As shown by the model Hamiltonian (\ref{H}), one has to prepare the localized impurity and the itinerant majorities in one dimension. Actually this has been successfully realized in a recent experiment of $^{173}$Yb atoms\cite{Ludwig-Maximilians-2018}.  Specifically, in this experiment, a  spin-independent deep lattice potential with a magic wavelength is imposed in the transverse direction to create a quasi-1D geometry, while in the longitudinal direction a spin-dependent lattice is imposed with a non-magic wavelength, such that the ac-polarizability of excited $P$-state is much larger than that of ground $S$-state. In this way, the $P$-state feels a much deeper longitudinal  confinement than $S$-state, and they can serve respectively as the localized impurity and itinerant fermions. In this work, we have simplified the longitudinal lattice potential felt by $S$-state as a harmonic trap, which should be a good approximation to the energy scale nearby the bottom of this potential.   

Another important ingredient to realize our model is to enhance the spin-exchange coupling via the confinement-induced-resonance(CIR)\cite{IASTU-2016,IASTU-2017,Zhang-Zhang-2018,Renmin-University,Zhang-Zhang-2020,Zhang-Cheng-Zhang-Zhai}. As shown experimentally\cite{Ludwig-Maximilians-2014,Fallani-2014,Kyoto}, the spin exchange coupling $J$ is determined by the difference of coupling strengths in the nuclear spin triplet and singlet channels, respectively denoted by $g_+$ and $g_-$. At the location of CIR, either $g_+$ or $g_-$ can be tuned to resonance, thereby maximizing the spin-exchange strength $|J|$. The enhancement of $J$ by CIR has been observed in ${}^{173}Yb$ system \cite{Ludwig-Maximilians-2018}. Given the relation\cite{Ludwig-Maximilians-2014,Fallani-2014,Kyoto}
\begin{equation}
J \propto \frac{g_- -g_+}{2} ,\quad U \propto \frac{g_++3g_-}{4},
\end{equation}
we can see that both $J$ and $U$ can go through resonance at the CIR of $g_+$ or $g_-$. 
Nearby the CIR of $g_+$ (or $g_-$), we have the ratio $U/J=-1/2$ (or $3/2$). For instance, at $U/J=-1/2$, from Eq.(\ref{eff_gamma}) we have $\gamma_s=-2J$ and $\gamma_t=0$, therefore a sequence of deep bound states only appear in $J>0$ side with energy dominated by the spin-singlet binding energy, while in $J<0$ side no deep bound states appear. In comparison, at $U/J=3/2$, we have $\gamma_s=0$ and $\gamma_t=2J$, thus all deep bound states appear in $J<0$ side but not $J>0$ side. Away from CIRs, the ratio $U/J$ is expected to be conveniently tuned to other values, such as the ones displayed in Fig.\ref{fig_U}.
In addition, since different alkali-earth atoms (such as ${}^{173}Yb$, ${}^{171}Yb$ and ${}^{87}Sr$) can have different $g_{\pm}$, we expect this diversity can provide an even rich tunability of $J$, $U$, and $U/J$ in realistic experiment. Given all of above, we expect the results discussed in Sec.\ref{sec3} and Sec.\ref{sec4} of this paper are highly relevant to the system of alkali-earth atoms in current experiment.

Finally, the energy spectrum in our work can be probed using the radio-frequency (rf) spectroscopy. For instance, by transferring the minority atoms from non-interacting to strongly interacting state in an inverse rf spectroscopy, both the attractive and repulsive branches of polarons have been successfully detected in $^{173}$Yb atoms\cite{PRL2019}. Besides, our work has concentrated on a small clusters of itinerant fermions, which requires a careful manipulation of particle numbers. This can be implemented by using the deformed potentials to spill atoms, as successfully demonstrated in cold atoms experiment\cite{Jochim}. By probing the energy spectra of the one-fermion and two-fermion systems, the Kondo screening effect can be tested by comparing the ground state energies of the two systems at both sides of $J>0$ and $J<0$, as shown in Fig.\ref{fig_3body}.

\section{Conclusion}  \label{sec6}

In this work, we have exactly solved few-body problem of one and two fermions in a 1D harmonic trap interacting with a local impurity for an arbitrary spin-exchange coupling strength $J$.    
It is found that the spin-exchange interaction can lead to a number of unique phenomena that cannot be achieved by pure contact interaction. These phenomena include the Kondo screening effect, the coexistence of FM upper branches and attractive lower branches under the AFM coupling, and the quadratic energy scaling in weak coupling regime. These unique properties can be extended to many-body system with spin-exchange interaction. Moreover, we also discuss the effect of an additional contact interaction, which effectively changes the parameter regime to support bound states with different magnetic structures. These unique features of spin-exchange interaction can be explored in the current experiments of ultracold alkaline-earth atoms.

\bigskip

{\bf Acknowledgement.} We thank Natan Andrei for helpful communication on the Bethe-Ansatz solution of Kondo model. The work is supported by the National Key Research and Development Program of China (2018YFA0307600, 2016YFA0300603), the National Natural Science Foundation of China (No.11534014), and the Strategic Priority Research Program of Chinese Academy of Sciences (No. XDB33000000).

\appendix

\section{Asymptotic energy expansion in strong coupling limit}

\subsection{One fermion case}\label{app_a1}
 
In the strong coupling regime, the upper-branch energies of one-fermion system take the form of:
\begin{equation}
    E_m \simeq  (2m+1)\omega - \frac{A_m}{\gamma},
\end{equation}
where $m$ is the level index, $\gamma$ is the effective coupling in the spin-singlet or triplet channel, and coefficient $A_m$ can be calculated as:
\begin{equation}
\begin{split}
    A_m &= \lim_{\gamma\to \infty} -\frac{\partial E(\gamma)}{\partial \frac{1}{\gamma} }\\
    \quad &= \lim_{\gamma\to \infty} \int dx \gamma\psi_\gamma(x)\delta(x)\gamma\psi_\gamma(x) \\
    \quad &= \lim_{\gamma\to \infty} \gamma\psi_\gamma(0)\cdot \gamma\psi_\gamma(0). 
    %\phi^{'}_{2m+1}(0)\cdot \phi^{'}_{2m+1}(0)\\
\end{split}
\end{equation}
By using the boundary condition $\gamma\psi_\gamma(0) = \frac{\psi'_\gamma(0)}{M}$, we can arrive at the expansion form of Eq.(\ref{asymptotic}) in the main text.

\subsection{Two fermions case} \label{app_a2}

We take the eigen-states saturating at $E=5\omega$ for example. In Fig.\ref{fig_3body}, we see that there are three upper branches saturating at $E=5\omega$ as $|J| \to \infty$. It can be shown that these upper branches can be divided into two classes. The first class of upper branch has coupling with the attractive lower branches, and therefore there will be an avoided level crossing when they meet in the spectrum. The second class of upper branch is orthogonal to all attractive lower branches, which leads to a direct level crossing when they meet. 

From the knowledge of one fermion spectrum, the lowest eigen-energies for the upper branches are $E=1.5,3.5\omega$ in the strong coupling regime $|J| = \infty$. When adding one more fermion, the strong fermion-impurity interaction requires the boundary condition : 
\begin{equation}
    \psi(x_1\sigma_1,x_2\sigma_2) = 0 \quad \text{if} \quad x_i =0
\end{equation}
Combining with fermion exchange anti-symmetry, we construct the following wave functions for  three degenerate eigen-states:
\begin{equation}
    \begin{split}
        &\left|\psi\right>^{(1)}_{deg} =  \frac{1}{\sqrt{2}} (\psi_1(x_1)\psi_3(x_2)- \psi_3(x_1)\psi_1(x_2)) \left|\uparrow \uparrow \Downarrow \right>\\
        &\left|\psi\right>^{(2)}_{deg} = \frac{1}{\sqrt{2}}\psi_1(x_1)\psi_3(x_2)\left|\uparrow \downarrow \Uparrow \right> - \frac{1}{\sqrt{2}}\psi_3(x_1)\psi_1(x_2)\left|\downarrow \uparrow \Uparrow \right>\\
        &\left|\psi\right>^{(3)}_{deg} = \frac{1}{\sqrt{2}}\psi_1(x_1)\psi_3(x_2)\left|\downarrow \uparrow \Uparrow \right> - \frac{1}{\sqrt{2}}\psi_3(x_1)\psi_1(x_2)\left|\uparrow \downarrow \Uparrow \right>\\
    \end{split}
\end{equation}
with $\psi_n(x) = \phi_n(x) {\rm sgn}(x)$ , where $\phi_n(x)$ is nth-eigenfunction of harmonic trap.

For large but finite coupling ($1/J \neq 0 $), the above degeneracy will be lifted and the three energy levels split as:
\begin{equation}
    E_m(J) = E_0 - \frac{\kappa_m }{J} \quad \text{with}  \quad m =1,2,3  \label{Em}
\end{equation}
%Here we begin to use an effective perturbation theory\cite{Yang-Cui} to calculate $\kappa_m$. First we assume three exact eigen wavefunction $\left|\psi^m_J\right>$corresponding $\frac{1}{J}$ nearby 0 with coefficients to be determined:
Accordingly, the zero-th order eigenstates can be a linear combination of $\left|\psi^{(m)}_{deg}\right>$ as:  
    \begin{equation}
        \left|\psi^{(m)}_{\infty} \right> = \sum_n a_{mn} \left|\psi^{(n)}_{deg}\right>, \label{eq1}
    \end{equation}
 which can be reorganized as
    \begin{equation}
        \begin{split}
            \left|\psi^{(m)}_J\right> &=  \psi^{(m)}_{J1}\left|\uparrow \uparrow \Downarrow \right> + \psi^{(m)}_{J2} \left|\uparrow \downarrow \Uparrow \right>  +\psi^{(m)}_{J3} \left|\downarrow \uparrow \Uparrow \right> \equiv\left(
            \begin{array}{c}
                \psi_{J1}^{(m)} \\
                \psi_{J2}^{(m)}\\
                \psi_{J3}^{(m)}
            \end{array}
            \right)\\
        \end{split} \label{eq2}
    \end{equation}    
Eqs.(\ref{eq1},\ref{eq2}) determine the relation between $\{\psi_{Jn}^{(m)}\} $ and $\{a_{mn}\}$. 

From the definition in Eq.(\ref{Em}), we have 
\begin{widetext}   
    \begin{equation}
        \begin{split}
            \kappa_m \delta_{mn} &= \lim_{J\to \infty} \left<\psi^{(m)}_J\right| \frac{\partial\hat{H}}{\partial(\frac{-1}{J})} \left|\psi^{(n)}_{J}\right>\\
                &= \lim_{J\to \infty} \left<\psi^{(m)}_J\right| 2J^2(\delta(x_1)\hat{S_1}\cdot \hat{S_0}+\delta(x_2)\hat{S_2}\cdot \hat{S_0}) \left|\psi^{(n)}_{J}\right>\\
                &= \lim_{J\to \infty} \int  dx_2 \Psi^{(m)\dagger}_J(0,x_2) 2J^2 \hat{S_1}\cdot \hat{S_0} \Psi^{(n)}_J(0,x_2) \\
                &\quad + \int dx_1 \Psi^{(m)\dagger}_J(x_1,0) 2J^2 \hat{S_2}\cdot \hat{S_0} \Psi^{(n)}_J(x_1,0) \\
        \end{split} \label{kappam}
    \end{equation}
\end{widetext}

Given the Schrodinger equation $\hat{H} \left|\psi^{(m)}_J\right> = E_m(J)\left|\psi^{(m)}_J\right>$,  we can integrate over $x_1, x_2$ separately and get the boundary conditions as:

    \begin{equation}
        \begin{split}
            &\frac{1}{2M}\partial_{x_1} \Psi^{(m)}_J(x_1,x_2)|_{x_1=0^-}^{x_1=0^+}= 2J \hat{S_1}\cdot \hat{S_0}\Psi^{(m)}_J(0,x_2) \\
            &\frac{1}{2M}\partial_{x_2} \Psi^{(m)}_J(x_1,x_2)|_{x_2=0^-}^{x_2=0^+}= 2J \hat{S_2}\cdot \hat{S_0} \Psi^{(m)}_J(x_1,0)
        \end{split}
    \end{equation}

Use above boundary conditions we can simplify (\ref{kappam}) as:    
\begin{widetext}
    \begin{equation}
        \begin{split}
            \kappa_m \delta_{mn} &=  (\frac{1}{2M})^2 \int dx_2 \partial_{x_1} \Psi^{m\dagger}_{\infty}(x_1,x_2) |^{x_1=0^+}_{x_1 = 0^-} \frac{1}{2\hat{S_1}\cdot \hat{S_0}} \partial_{x_1} \Psi^{n}_{\infty}(x_1,x_2) |^{x_1=0^+}_{x_1 = 0^-}\\
            & \quad +  (\frac{1}{2M})^2 \int dx_1 \partial_{x_2} \Psi^{m\dagger}_{\infty}(x_1,x_2) |^{x_2=0^+}_{x_2 = 0^-} \frac{1}{2\hat{S_1}\cdot \hat{S_0}} \partial_{x_2} \Psi^{n}_{\infty}(x_1,x_2) |^{x_2=0^+}_{x_2 = 0^-}\\
           &= \frac{1}{M^2}(a_{m1},a_{m2},a_{m3}) \int dx
            \left(
            \begin{array}{c}
                \frac{4(B-A)^2}{3} \quad \frac{8B(B-A)}{3}  \quad \frac{8A(A-B)}{3}\\
                \frac{8B(B-A)}{3} \quad 4 A^2+\frac{4 B^2}{3}  \quad -\frac{16AB}{3} \\
                \frac{8A(A-B)}{3} \quad  -\frac{16AB}{3} \quad \frac{4 A^2}{3}+4 B^2
            \end{array}
            \right) 
            \left(
            \begin{array}{c}
                a_{n1}\\
                a_{n2}\\
                a_{n3}\\
            \end{array}
            \right)\\
            & =  \omega \sqrt{\frac{\omega}{M}}\frac{1}{\sqrt{\pi}}(a_{m1},a_{m2},a_{m3})
            \left(
            \begin{array}{c}
                \frac{10}{3} \quad 4  \quad \frac{8}{3}\\
                4 \quad 6 \quad 0 \\
                \frac{8}{3} \quad 0 \quad \frac{22}{3}
            \end{array}
            \right) 
            \left(
            \begin{array}{c}
                a_{n1}\\
                a_{n2}\\
                a_{n3}\\
            \end{array}
            \right)\\    
        \end{split}
    \end{equation} 
\end{widetext}
where we have the notions $A = \frac{1}{\sqrt{2}}\phi_1'(0)\psi_3(x) , B =\frac{1}{\sqrt{2}} \phi_3'(0)\psi_1(x) $, and in the last step we have used $\int dx A^2 = \frac{(M\omega)^\frac{3}{2}}{\sqrt{\pi}} $ , $\int dx B^2 = \frac{3 (M\omega)^\frac{3}{2} }{2\sqrt{\pi}} $  and $\int dx AB = 0$. 

Finally we diagonalize the matrix and obtain the eigen vector of $a_{mn}$(without normalization temporarily) :
\begin{equation}
        \left(
            \begin{array}{c}
            1\\
            1\\
            1\\
            \end{array}
        \right),
        \left(
            \begin{array}{c}
            \frac{1}{2} \left(\sqrt{7}-3\right)\\
            \frac{1}{2} \left(1-\sqrt{7}\right)\\
            1\\
            \end{array}
        \right),
        \left(
            \begin{array}{c}
            \frac{1}{2} \left(-\sqrt{7}-3\right)\\
            \frac{1}{2} \left(\sqrt{7}+1\right)\\
            1\\
            \end{array}
        \right)\\
\end{equation}
Note that the first state is the full FM state. The corresponding coefficients for the three states are (in unit of $ \omega \sqrt{\frac{\omega}{M}}$):
\begin{equation}
\kappa_m= \frac{10}{\sqrt{\pi}},\frac{2}{3\sqrt{\pi}} \left(2 \sqrt{7}+5\right),\frac{2}{3\sqrt{\pi}} \left(5-2 \sqrt{7}\right) 
\end{equation}
We have checked that the coefficients are consistent with our numerical fitting of energy spectrum in strong coupling regime.

\section{Formula of few-body problem with spin-exchange interaction and an additional contract potential}\label{app_b}

Here we take the two-fermion case for instance. After adding the contact interaction term $U\delta(x)$, Eq.(\ref{eq_3b}) in the main text then changes to :
\begin{widetext}
    \begin{equation}
        \begin{split}
        &\phi^{(1)}_{mn} = \frac{1}{E-E_m-E_n} \cdot \frac{1}{2} \cdot  \sum_{p}-V_{mp}\phi^{(2)}_{np}+V_{np}\phi^{(2)}_{mp} +V_{np}\phi^{(1)}_{pm}- V_{mp}\phi^{(1)}_{pn} + 2U_{mp}\phi^{(1)}_{pn}-2U_{np}\phi^{(1)}_{pm} \\
        &\phi^{(2)}_{mn} = \frac{1}{E-E_m-E_n} \sum_p V_{np}\phi^{(1)}_{mp}-V_{np}\phi^{(1)}_{pm}+ \frac{1}{2} V_{mp}\phi^{(2)}_{pn}- \frac{1}{2}V_{np}\phi^{(2)}_{mp} + U_{mp}\phi^{(2)}_{pn} + U_{np}\phi^{(2)}_{mp}\\
        \end{split}
    \end{equation}
\end{widetext} 
With the same definition of $\tilde{F}^{(i)}$ in the main text, we obtain the following matrix equation:
\begin{widetext}
     \begin{equation}
      \left(
        \begin{array}{ccc}
        \frac{1}{4} (e-2 q) (3 -2 \frac{U}{J}) & \frac{3}{4} e (1+2 \frac{U}{J}) & \frac{1}{4} (-3) e (1+2 \frac{U}{J}) \\
        e \left(\frac{\frac{U}{J}}{6}-\frac{1}{4}\right) & -\frac{1}{4} (e-2 q) (1+2 \frac{U}{J}) & -\frac{1}{4} e (1+2 \frac{U}{J}) \\
        \frac{1}{6} e (3 -2 \frac{U}{J}) & -\frac{1}{2} e (1+2 \frac{U}{J}) & \frac{1}{2} q (1+2 \frac{U}{J}) \\
        \end{array}
      \right)
        \left(
            \begin{array}{c}
                \tilde{F}^{(1)} \\
                \tilde{F}^{(2)} \\
                \tilde{F}^{(3)} \\
            \end{array}
        \right)
        =\frac{1}{J}
        \left(
            \begin{array}{c}
                \tilde{F}^{(1)} \\
                \tilde{F}^{(2)} \\
                \tilde{F}^{(3)} \\
            \end{array}
        \right)  
    \end{equation}
\end{widetext}
with $U_{mn} = U\phi_m\phi_n$, which can be diagonalized to obtain the spectrum for various ratios $\frac{U}{J}$.

\section{Bethe ansatz solution with linear fermion dispersion} \label{app_c}

By assuming a linear dispersion of itinerant fermions $\epsilon_k\sim k$, Ref.\cite{Andrei-Furuya-Lowenstein} has provided Bethe ansatz solutions for the Kondo problem in 1D. From equations (2.47, 2.48, 3.12) therein, one can obtain the ground state energy as:
\begin{eqnarray}
&& E_{gs}(J)-E_{gs}(J=0) \nonumber\\
&=&D\cdot\left( \int\frac{1}{2c}\frac{\Theta(2\Lambda-2)-\pi}{\cosh\frac{\pi}{c}\Lambda}d\Lambda +i\ln\frac{1-\frac{iJ}{2}}{1+\frac{iJ}{2}} \right)
\end{eqnarray}
where $\Theta(x) = -2\arctan(\frac{x}{c})$ , $c = \frac{2J}{1-\frac{3J^2}{4}}$ and $D$ is electron number density. Note that here $J$ is scaled by $k_F/M$ ($k_F$ is the Fermi momentum of itinerant fermions) and becomes dimensionless. This result shows that the leading order of ground state energy $E_{gs}(J)-E_{gs}(J=0) \propto J^3$ nearby $J\sim 0$.

\section{Bound state of two untrapped fermions interacting with a localized impurity} \label{app_d}

Here we consider the system of two untrapped fermions and one impurity interacting with spin-exchange interaction. In this case our matrix equation Eq.(\ref{final_eq}) is still valid except that the harmonic basis should be replaced by momentum basis. For instance, one of the replacement is $\phi_m(0) \to \phi_k(0)= \frac{1}{\sqrt{L}}$, where $L$ is the length of 1D system. Accordingly, the matrix elements in momentum space are $e_{kk'} = \frac{1}{L}\frac{1}{E-E_k-E_{k'}}$ , $q_{kk'} = \delta_{kk'}  \frac{1}{L}\sum_p \frac{1}{E-E_k-E_p}$.

To find the bound state solution, we numerically solve Eq.(\ref{final_eq}) in momentum space by discretizing the integral equation using a Gauss-Legendre type and calculating the determinant of the corresponding linear equation. With a cutoff momentum $k_{\Lambda}/(MJ)=50$ and $2000$ Gauss points, error can be controlled in $10^{-3}$ .  At $J<0$ we obtain the lowest solution $E\approx -0.151J^2$, which means both fermions are bound with the impurity. Above this ground state, continuous solutions start from $-\frac{J^2}{8}$ , which means only one tightly bound fermion with the impurity to form a spin triplet. While at $J>0$ there are only  continuous solutions starting from $-\frac{9J^2}{8}$ which means only one tightly bound fermion with the impurity to form a spin singlet. This picture of Kondo screening effect is consistent with strong coupling limit of our trapped system, as discussed in the main text.

\end{document}